\documentstyle[prb,aps,multicol,epsf]{revtex}

\begin{document}

\draft
\title{
Resonant tunneling through a quantum dot
weakly coupled to quantum wires or quantum Hall edge states
}
\author{A. Furusaki}
\address{Yukawa Institute for Theoretical Physics, Kyoto University,
Kyoto 606-01, Japan}
\maketitle
\begin{abstract}
Resonant tunneling through a quantum dot weakly coupled to
Tomonaga-Luttinger liquids is discussed.
The linear conductance due to sequential tunneling is calculated by
solving a master equation for temperatures below and above the average
level spacing in the dot.
When the parameter $g$ characterizing the Tomonaga-Luttinger liquid is 
smaller than 1/2, the resonant tunneling process is incoherent down to 
zero temperature.
At low temperature $T$ the height and width of the conductance peaks
in the Coulomb blockade oscillations are proportional to
$T^{\frac{1}{g}-2}$ and $T$, respectively.
The contribution from tunneling via a virtual intermediate state
(cotunneling) is also included.
The resulting conductance formula can be applied for the resonant
tunneling between edge states of fractional quantum Hall liquids
with filling factor $\nu=1/(2m+1)=g$.
\end{abstract}

\begin{multicols}{2}

\section{Introduction}
\label{sec:intro}

Advances in nanostructure technology have made it possible to
fabricate semiconductor devices such as quantum dots,\cite{Kastner}
small two-dimensional regions in which electrons are confined.
The transport properties of the quantum dots weakly coupled via tunnel
barriers to external leads have recently attracted much attention.
At low temperatures the linear conductance exhibits
periodic peak structures as a function of a gate voltage, a phenomenon 
known as Coulomb blockade oscillations.
These peaks occur when the energy change due to tunneling of one
electron into or out of the dot equals the Fermi energy of the leads.
Apart from these resonance points, tunneling is suppressed due to the
Coulomb blockade.\cite{AverinLikharev,Grabert}
Single-particle energy levels in a small quantum dot are discrete with
mean level spacing $\Delta$ and have decay width $\Gamma_{L(R)}$ which
is proportional to the tunneling rate to the left (right) lead. 
In the temperature regime $\Gamma_{L(R)}\ll T\ll\Delta$, the line
shape of the conductance peaks is\cite{Beenakker}
\begin{equation}
G=
\frac{e^2}{\hbar}\frac{\Gamma_L\Gamma_R}{\Gamma_L+\Gamma_R}
\left(-\frac{d}{d\varepsilon}\right)\frac{1}{e^{\varepsilon/T}+1},
\label{G_Fermi}
\end{equation}
where $\varepsilon$ is proportional to the gate voltage measured
relative to the resonance point.
Equation (\ref{G_Fermi}) is valid when the external leads are Fermi
liquids and has been used to interpret numerous experimental data.

\begin{figure}
\narrowtext
\begin{center}\leavevmode\epsfxsize=65mm \epsfbox{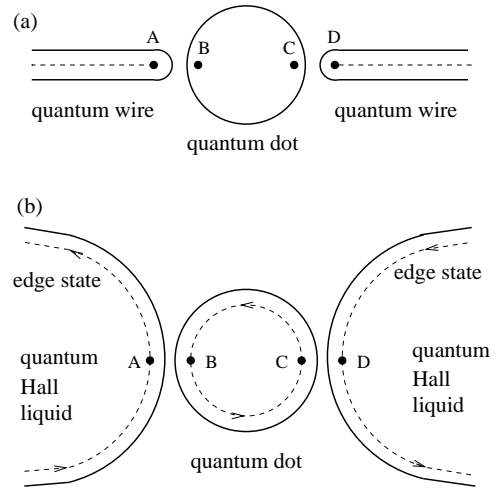}\end{center}
\caption{Quantum dot coupled to (a) quantum wires and (b) edge states
in quantum Hall liquids.   Dotted lines represent one-dimensional
states.} 
\label{fig:dots}
\end{figure}

In this paper we discuss a generalization of Eq.~(\ref{G_Fermi}) to
the case where the external leads are Tomonaga-Luttinger (TL)
liquids. 
The TL liquids can be realized in very narrow quantum wires
\cite{Tarucha} or as edge states of fractional quantum Hall
liquids.\cite{Wen,Milliken,Chang} 
Figure \ref{fig:dots} shows schematic pictures of the systems of our
interest, a quantum dot coupled via tunnel barriers to one-dimensional 
(1D) quantum wires or to edge states in fractional quantum Hall
liquids. 
In both cases it is assumed that there are small but finite matrix
elements for the tunneling from the points B and C of the
zero-dimensional (0D) states formed in the quantum dot to the points A
and D of the 1D TL liquids.
To fully understand the transport in these systems,
it is necessary to take into account both the charging energy and the
discrete energy levels in the dot.

This paper is intended to give a simplified description of the
resonant tunneling between TL liquids.
We extend the theory of the sequential tunneling developed in
Refs.~\onlinecite{FuruNaga} and \onlinecite{Chamon} to derive a
generalized formula of the linear conductance, and discuss the
validity of the sequential-tunneling picture.
Instead of applying the instanton technique starting from an effective 
action for bosonic variables,\cite{FuruNaga} we use a master-equation
approach to calculate the resonant current through a quantum
dot.\cite{note_Chamon}
This method is more direct and transparent, and has been successful in 
describing the Coulomb blockade oscillations in quantum
dots coupled to Fermi-liquid leads.\cite{Beenakker,GlazShek}
With this method we can describe in a unified way the resonant
tunneling through a quantum dot which is weakly coupled to TL-liquid
reservoirs as well as to ordinary Fermi-liquid reservoirs.
We will see that our formula has even the same form as the conductance 
of double quantum dots weakly coupled to Fermi-liquid
reservoirs.\cite{MatvGlazBara} 
Another merit of this approach is that it allows easily to treat a
system with two TL-liquid reservoirs having different interaction
parameters.
This situation corresponds to, for example, the resonant tunneling
from a $\nu=1/3$ edge state to a $\nu=1$ edge state.

The use of the master-equation approach can be justified because the
resonant current is mainly carried by sequential tunneling processes
down to zero temperature when the interaction parameter $g$
characterizing TL liquids is smaller than 1/2.\cite{FuruNaga}
When $g\ge1/2$, on the other hand, there is a crossover temperature
below which the transport becomes coherent and the line shape of the
peak conductance approaches the
universal form obtained by Kane and Fisher.\cite{KaneFish}
For the edge states in the fractional quantum Hall liquid with a
filling factor $\nu=1/(2m+1)$, the parameter $g$ equals
$\nu$.\cite{Wen}
This means that the tunneling between edge states via a quantum dot
can be described, down to zero temperature, by the
sequential-tunneling picture discussed in this paper, unless the bare
tunneling elements are large and/or $\nu=1$. 
Finally we note that our approach is different from the previous work
by Kinaret {\it et al.}\cite{Kinaret} in which the edge states in the
quantum dot are described as chiral TL liquids but the leads are
assumed to be Fermi liquids.
In our study, on the other hand, the leads are described as TL liquids 
and the states in the dot are treated as 0D states.
Thus we are interested in the effects coming from anomalous
power-law correlations in the TL leads.

\section{Model}
\label{sec:model}

In this section we introduce a simple model for a dot and leads, and
calculate propagators in the leads.
The Hamiltonian of the system shown in Fig.~\ref{fig:dots} can be
separated into four parts, $H=H_L+H_R+H_D+H_T$, with $H_{L(R)}$
describing the left (right) lead, $H_D$ the dot, and $H_T$ the
tunneling between the leads and the dot.
Since the tunneling rate through the tunnel barriers is assumed to be
very small, the number of electrons in the dot is a good quantum
number.
Thus we may write the dot Hamiltonian
\begin{equation}
H_D=\sum_N\sum^\infty_{i=0} E(N,i)|N,i\rangle\langle N,i|,
\label{H_D}
\end{equation}
where $N$ is the number of electrons in the dot and $E(N,i)$ is
eigenenergy of the many-body state $|N,i\rangle$
[$E(N,0)<E(N,1)<E(N,2)<\ldots$]. 
In this paper we will not try to calculate the energies $E(N,i)$
themselves, but instead we assume that they are given, since
we are mainly interested in the line shape of conductance peaks in the
Coulomb blockade oscillations, not in the position of these peaks.
Without loss of generality, we may assume that $E(N_0,0)$ and
$E(N_0+1,0)$ are the two lowest energies among $E(N,i=0)$'s.
The energy differences $E(N_0-1,0)-E(N_0,0)$ and
$E(N_0+2,0)-E(N_0+1,0)$  
are of the order of the charging energy $E_C$.
A convenient choice to represent this would be the approximation
usually made in studies of the Coulomb blockade:\cite{AverinLikharev}
$E(N,0)=(E_C/2)(N-{\cal N})^2$, where ${\cal N}$ ($N_0<{\cal
N}<N_0+1$) is an external parameter which can be controlled by
changing the gate voltage. 
Since we are interested in the temperature range $T\ll E_C$,
we neglect the states in which $N\ne N_0,N_0+1$.
Then the current flow is accompanied by the periodic change of the
electron numbers,\cite{GlazShek}
$N_0\to N_0+1\to N_0\to\ldots$.
We also note that the ratio $\Delta/E_C$ can be much smaller than 1
for both systems shown in Fig.~\ref{fig:dots}.\cite{FuruMatv,Yi}

First we address the situation shown in Fig.~\ref{fig:dots}(a).
The left (L) and right (R) leads are described as TL liquids whose
interaction parameter is $K_{\rho L(R)}$
for the charge sector and $K_{\sigma L(R)}$ for the spin
sector.\cite{Schulz}
Due to repulsive electron-electron interactions $K_\rho$ is less than
1 while $K_\sigma$ is fixed at 1 because of the SU(2) spin symmetry.
In the bosonized form $H_L$ is written as
\begin{equation}
H_L=
\hbar\int^\infty_0dk
\left(v^{}_{cL}ka^\dagger_{k,L}a^{}_{k,L}
+v^{}_{sL}kb^\dagger_{k,L}b^{}_{k,L}\right),
\label{H_L}
\end{equation}
where $a_{k,L}$ ($b_{k,L}$) is an annihilation operator of bosons
describing charge (spin) density fluctuations propagating with velocity
$v^{}_{c(s)L}$.
The Hamiltonian of the right lead, $H_R$, can be written in a similar
way. 
The tunneling Hamiltonian is given by
\begin{eqnarray}
H_T&=&
t_L\sum_\sigma\left[
\psi^\dagger_\sigma({\rm A})\psi^{}_\sigma({\rm B})
+\psi^\dagger_\sigma({\rm B})\psi^{}_\sigma({\rm A})
\right]
\nonumber\\&&
+
t_R\sum_\sigma\left[
\psi^\dagger_\sigma({\rm C})\psi_\sigma({\rm D})
+\psi^\dagger_\sigma({\rm D})\psi_\sigma({\rm C})
\right],
\end{eqnarray}
where $\psi_\sigma({\rm X})$ annihilates an electron with spin up
($\sigma=+1$) or down ($\sigma=-1$) at the point X
(X=A,B,C, and D in Fig.~\ref{fig:dots}).
The electron field operator at the boundary may be written
as\cite{Eggert}
\end{multicols}\widetext
\vspace{-6mm}\noindent\underline{\hspace{90mm}}
\begin{equation}
\psi_\sigma({\rm A})=
\sqrt{\frac{2}{\pi\alpha}}\exp\left[
\int^\infty_0dk\frac{e^{-\alpha k/2}}{\sqrt{2K_{\rho L}k}}
\left(a^{}_{k,L}-a^\dagger_{k,L}\right)
+\sigma\int^\infty_0dk\frac{e^{-\alpha k/2}}{\sqrt{2k}}
\left(b^{}_{k,L}-b^\dagger_{k,L}\right)
\right],
\label{psi}
\end{equation}
where $\alpha$ is a short-distance cutoff of the order of the
reciprocal of the Fermi wave number $k_F$.
This leads to local propagators
\begin{equation}
\langle\psi^\dagger_\sigma({\rm A},t)\psi^{}_\sigma({\rm A},0)\rangle_L
=
\langle\psi^{}_\sigma({\rm A},t)\psi^\dagger_\sigma({\rm A},0)\rangle_L
=
\frac{c^{}_A}{\alpha}
\left\{
\frac{i\Lambda}{\pi T}\sinh\!\left[\frac{\pi T(t-i\delta)}{\hbar}\right]
\right\}^{-\frac{1}{g^{}_L}},
\label{G_A}
\end{equation}
where $c^{}_A$ is a dimensionless constant of order 1, $\Lambda$ is a
high-energy cutoff or a band width, 
$\delta$ is positive infinitesimal, and
$g^{-1}_L=\frac{1}{2}\left(\frac{1}{K_{\rho L}}+1\right)$. 
The thermal averages are calculated with respect to $H_L$, and
$\psi^{}_\sigma({\rm A},t)=
\exp(iH_Lt/\hbar)\psi^{}_\sigma({\rm A})\exp(-iH_Lt/\hbar)$.
Similarly, the propagators at the point D are obtained as
\begin{equation}
\langle
\psi^\dagger_\sigma({\rm D},t)\psi^{}_\sigma({\rm D},0)
\rangle_R
=
\langle
\psi^{}_\sigma({\rm D},t)\psi^\dagger_\sigma({\rm D},0)
\rangle_R
=
\frac{c^{}_D}{\alpha}
\left\{\frac{i\Lambda}{\pi T}
\sinh\!\left[\frac{\pi T(t-i\delta)}{\hbar}\right]
\right\}^{-\frac{1}{g^{}_R}},
\label{G_D}
\end{equation}
\begin{multicols}{2}\noindent
where $c^{}_D$ is a dimensionless constant,
$g^{-1}_R=\frac{1}{2}\left(\frac{1}{K_{\rho R}}+1\right)$
and the averages are taken with respect to $H_R$.
Note that $g^{}_L$ and $g^{}_R$ are smaller than 1 because
both $K_{\rho L}$ and $K_{\rho R}$ are smaller than 1.

Although we have considered the system shown in
Fig.~\ref{fig:dots}(a), both Eqs.~(\ref{G_A}) and (\ref{G_D}) also
hold for the quantum Hall edge states in
Fig.~\ref{fig:dots}(b).
Suppose that the left lead and the right lead are in the
quantum Hall regime with filling factors $\nu^{}_L=1/(2l+1)$
and $\nu^{}_R=1/(2m+1)$ ($l,m$: integers),
respectively.\cite{note1} 
In this case the Hamiltonian for the left edge states is
\begin{equation}
H_L=\frac{v}{4\pi}\int dx\left(\frac{d\varphi(x)}{dx}\right)^2,
\label{H_L-2}
\end{equation}
where $x$ is the coordinate along the edge and the bosonic field
$\varphi(x)$ obeys $[\varphi(x),\varphi(y)]=i\pi{\rm
sgn}(x-y)$.\cite{Wen} 
Since the electron field operator at the point A is given by
$\psi({\rm A},t)\propto\exp[i\varphi({\rm
A},t)/\sqrt{\nu^{}_L}]$,\cite{Wen,note} 
we find the correlation function
in the left edge is also given by Eq.~(\ref{G_A}) with
$g^{}_L=\nu^{}_L$.
The same derivation also holds for the right edge:
$g^{}_R=\nu^{}_R$.
In the following sections, we shall use the parameters
$g^{}_L$ and $g^{}_R$ without distinguishing the two systems.
Since the electron spin is not important in the following discussion,
we will suppress the spin indices
in the electron field operator.

\section{Low-temperature regime ($\Gamma\ll T<\Delta$)} 
\label{sec:low-T}

\begin{figure}
\begin{center}
\leavevmode\epsfxsize=65mm \epsfbox{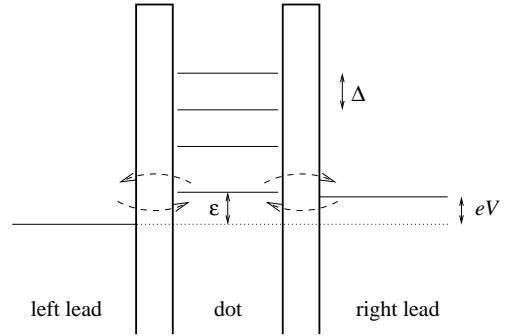}
\end{center}
\narrowtext
\caption{Schematic picture of energy diagrams. Horizontal lines in the 
leads represent Fermi levels.  There are four tunneling processes.}
\label{fig:processes}
\end{figure}

In this section we calculate the linear conductance for
$T\lesssim\Delta$ within the master-equation approach.\cite{GlazShek}
In this approach we assume that the energy is conserved in each
tunneling process, and neglect the contributions from tunneling via
virtual intermediate states.
We will see later that this assumption is valid near the conductance
peaks in the weak-tunneling limit.
In the following calculation we include only low-energy states
$|N,i\rangle$ with
$N=N_0$ or $N_0+1$ which are major contributors in
the conduction process at temperature $T\lesssim\Delta\ll E_C$, a
situation often satisfied in experiments using semiconductor quantum
dots.\cite{AverinKorotkov,Johnson} 
A schematic energy diagram is shown in Fig.~\ref{fig:processes}.

In lowest order in $H_T$ the transition rates from the state
$|N_0,i\rangle$ to the state $|N_0+1,j\rangle$ due to the tunneling of 
an electron into the dot through the left or right tunnel barrier are
calculated from the golden rule:
\end{multicols}\widetext
\vspace{-6mm}\noindent\underline{\hspace{90mm}}
\begin{mathletters}
\begin{eqnarray}
P_L(N_0,i;N_0+1,j)&=&
\left(\frac{t_L}{\hbar}\right)^2
\int^\infty_{-\infty}\!\!dt\,
e^{-i\varepsilon^{}_{ij}t/\hbar}
\left|
\langle N_0+1,j|\psi^\dagger({\rm B})|N_0,i\rangle
\right|^2
\langle
\psi^\dagger({\rm A},t)\psi({\rm A},0)
\rangle_L,
\label{P_L(N_0toN_0+1)}\\
P_R(N_0,i;N_0+1,j)&=&
\left(\frac{t_R}{\hbar}\right)^2
\int^\infty_{-\infty}\!\!dt\,
e^{-i(\varepsilon^{}_{ij}-eV)t/\hbar}
\left|
\langle N_0+1,j|\psi^\dagger({\rm C})|N_0,i\rangle
\right|^2
\langle
\psi^\dagger({\rm D},t)\psi({\rm D},0)
\rangle_R,
\label{P_R(N_0toN_0+1)}
\end{eqnarray}
where $\varepsilon^{}_{ij}=E(N_0+1,j)-E(N_0,i)$ and $eV$ is the
difference between the chemical potentials
of the left and the right leads.
The transition rates for the inverse processes are similarly given by
\begin{eqnarray}
P_L(N_0+1,j;N_0,i)&=&
\left(\frac{t_L}{\hbar}\right)^2
\int^\infty_{-\infty}\!\!dt\,
e^{i\varepsilon^{}_{ij}t/\hbar}
\left|
\langle N_0,i|\psi({\rm B})|N_0+1,j\rangle
\right|^2
\langle
\psi({\rm A},t)\psi^\dagger({\rm A},0)
\rangle_L,
\label{P_L(N_0+1toN_0)}\\
P_R(N_0+1,j;N_0,i)&=&
\left(\frac{t_R}{\hbar}\right)^2
\int^\infty_{-\infty}\!\!dt\,
e^{i(\varepsilon^{}_{ij}-eV)t/\hbar}
\left|
\langle N_0,i|\psi({\rm C})|N_0+1,j\rangle
\right|^2
\langle
\psi({\rm D},t)\psi^\dagger({\rm D},0)
\rangle_R.
\label{P_R(N_0+1toN_0)}
\end{eqnarray}
\end{mathletters}\noindent
Using the propagators (\ref{G_A}) and (\ref{G_D}), we evaluate the
integrals (\ref{P_L(N_0toN_0+1)})--(\ref{P_R(N_0+1toN_0)}),
\begin{mathletters}
\begin{eqnarray}
P_L(N_0,i;N_0+1,j)&=&
\frac{T}{\hbar}e^{-\varepsilon^{}_{ij}/2T}
\gamma^{}_L(N_0,i;N_0+1,j),
\label{P_L(N_0toN_0+1)-2}\\
P_R(N_0,i;N_0+1,j)&=&
\frac{T}{\hbar}e^{-(\varepsilon^{}_{ij}-eV)/2T}
\gamma^{}_R(N_0,i;N_0+1,j),
\label{P_R(N_0toN_0+1)-2}\\
P_L(N_0+1,j;N_0,i)&=&
\frac{T}{\hbar}e^{\varepsilon^{}_{ij}/2T}
\gamma^{}_L(N_0,i;N_0+1,j),
\label{P_L(N_0+1toN_0)-2}\\
P_R(N_0+1,j;N_0,i)&=&
\frac{T}{\hbar}e^{(\varepsilon^{}_{ij}-eV)/2T}
\gamma^{}_R(N_0,i;N_0+1,j),
\label{P_R(N_0+1toN_0)-2}
\end{eqnarray}
\end{mathletters}\noindent
where $\gamma^{}_L(N_0,i;N_0+1,j)$ and $\gamma^{}_R(N_0,i;N_0+1,j)$
are given by
\begin{mathletters}
\begin{eqnarray}
\gamma^{}_L(N_0,i;N_0+1,j)&=&
\frac{\Gamma_{Lij}}{2\pi T}
\left(\frac{\pi T}{\Lambda}\right)^{\frac{1}{g_L}-1}
\frac{\left|\Gamma\!\left(
            \frac{1}{2g_L}+i\frac{\varepsilon^{}_{ij}}{2\pi T}
             \right)\right|^2}
     {\Gamma\!\left(\frac{1}{g_L}\right)},
\label{gamma_L(N_0,i;N_0+1,j)}\\
\gamma^{}_R(N_0,i;N_0+1,j)&=&
\frac{\Gamma_{Rij}}{2\pi T}
\left(\frac{\pi T}{\Lambda}\right)^{\frac{1}{g_R}-1}
\frac{\left|\Gamma\!\left(
            \frac{1}{2g_R}+i\frac{\varepsilon^{}_{ij}-eV}{2\pi T}
             \right)\right|^2}
     {\Gamma\!\left(\frac{1}{g_R}\right)}.
\label{gamma_R(N_0,i;N_0+1,j)}
\end{eqnarray}
\end{mathletters}\noindent
We have defined
$\Gamma_{Lij}=(2\pi c^{}_At^2_L/\alpha\Lambda)
 |\langle N_0,i|\psi({\rm B})|N_0+1,j\rangle|^2$
and
$\Gamma_{Rij}=(2\pi c^{}_Dt^2_R/\alpha\Lambda)
 |\langle N_0,i|\psi({\rm C})|N_0+1,j\rangle|^2$.
The time evolution of $P(N_0,i)$, the probability that the state
$|N_0,i\rangle$ is occupied, obeys the master equation,
\begin{eqnarray}
\frac{\partial}{\partial t}P(N_0,i)&=&
\sum_j
\biggl\{
P(N_0+1,j)[P_L(N_0+1,j;N_0,i)+P_R(N_0+1,j;N_0,i)]
\nonumber\\&&\qquad\qquad
-P(N_0,i)[P_L(N_0,i;N_0+1,j)+P_R(N_0,i;N_0+1,j)]
\biggr\}.
\label{master_eq}
\end{eqnarray}
In the steady state where $(\partial/\partial t)P(N_0,i)=0$, we need
to solve a set of detailed balance relations:
\begin{eqnarray}
&&
P(N_0,i)
[P_L(N_0,i;N_0+1,j)+P_R(N_0,i;N_0+1,j)]
\nonumber\\&&\qquad
=P(N_0+1,j)
[P_L(N_0+1,j;N_0,i)+P_R(N_0+1,j;N_0,i)].
\label{detailed_balance}
\end{eqnarray}
In the equilibrium where $V=0$, $P(N_0,i)$ is given by
\begin{equation}
P_{\rm eq}(N_0,i)=\frac{\exp[-E(N_0,i)/T]}
                       {\sum_{N_0}\sum_i\exp[-E(N_0,i)/T]},
\label{P_eq}
\end{equation}
so that from Eqs.~(\ref{P_L(N_0toN_0+1)-2})--(\ref{P_R(N_0+1toN_0)-2})
we can easily see that Eq.~(\ref{detailed_balance}) is satisfied.
Following Ref.~\onlinecite{Beenakker}, we solve
Eq.~(\ref{detailed_balance}) to the first order in $V$.
We first substitute $P(N_0,i)=P_{\rm eq}(N_0,i)[1+\frac{eV}{T}p(N_0,i)]$
into Eq.~(\ref{detailed_balance}) and linearize it with respect to $V$.
We find
\begin{equation}
p(N_0+1,j)-p(N_0,i)=
\frac{\gamma^{}_R(N_0,i;N_0+1,j)}
     {\gamma^{}_L(N_0,i;N_0+1,j)+\gamma^{}_R(N_0,i;N_0+1,j)}.
\label{p}
\end{equation}
The current through the quantum dot is then given, up to first order
in $V$, by
\begin{eqnarray}
I&=&
-e\sum_{i,j}
[P(N_0,i)P_L(N_0,i;N_0+1,j)
 -P(N_0+1,j)P_L(N_0+1,j;N_0,i)]\nonumber\\
&=&
\frac{e^2V}{\hbar}\sum_{i,j}e^{-\varepsilon^{}_{ij}/2T}P_{\rm eq}(N_0,i)
\gamma^{}_L(N_0,i;N_0+1,j)
[p(N_0+1,j)-p(N_0,i)]
\label{I}
\end{eqnarray}
from which we get the linear conductance
\begin{equation}
G=
\frac{e^2}{\hbar}\sum_{i,j}e^{-\varepsilon^{}_{ij}/2T}P_{\rm eq}(N_0,i)
\frac{\gamma^{}_L(N_0,i;N_0+1,j)\gamma^{}_R(N_0,i;N_0+1,j)}
     {\gamma^{}_L(N_0,i;N_0+1,j)+\gamma^{}_R(N_0,i;N_0+1,j)},
\label{G}
\end{equation}
\begin{multicols}{2}\noindent
which is a generalization of Eq.~(3.14) of
Ref.~\onlinecite{Beenakker}.

\begin{figure}
\begin{center}\leavevmode\epsfxsize=65mm \epsfbox{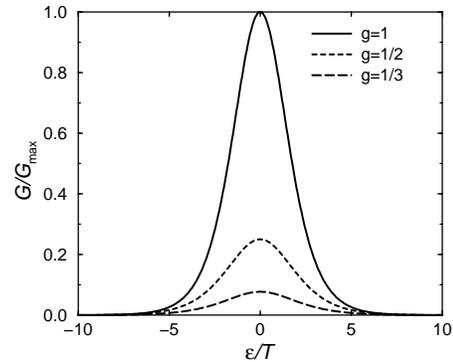}\end{center}
\narrowtext
\caption{The linear conductance due to the sequential tunneling,
Eq.~(\ref{G_sequential})
for $g^{}_L=g^{}_R=g=1,1/2,$ and $1/3$.  The conductance is normalized 
by the peak value of the $g=1$ case, $G_{\max}=e^2\Gamma/8\hbar T$.
We have set $T/\Lambda=0.25$.} 
\label{fig:g_s}
\end{figure}

For the rest of this section, let us concentrate on the temperature
regime $T\ll\Delta$. 
In this regime we may assume that the dot is always in the lowest
energy state with a given electron number, so that we may set $i=j=0$
in Eq.~(\ref{G}). 
We can thus write $P_{\rm eq}(N_0,0)$ as
\begin{equation}
P_{\rm eq}(N_0,0)=\frac{1}{1+e^{-\varepsilon/T}},
\label{P(N_0,0)}
\end{equation}
where $\varepsilon=E(N_0+1,0)-E(N_0,0)$.
From Eqs.~(\ref{G}) and (\ref{P(N_0,0)}) we find the linear
conductance\cite{note2} for $T\ll\Delta$,
\begin{equation}
G=
\frac{e^2}{2\hbar\cosh(\varepsilon/2T)}
\frac{\gamma^{}_L(\varepsilon,T)\gamma^{}_R(\varepsilon,T)}
     {\gamma^{}_L(\varepsilon,T)+\gamma^{}_R(\varepsilon,T)},
\label{G_sequential}
\end{equation}
where $\gamma^{}_{L,R}(\varepsilon,T)=\gamma^{}_{L,R}(N_0,0;N_0+1,0)$.
This is the generalization of Eq.~(\ref{G_Fermi}) to the case where
the leads are TL liquids.
When $g_{L(R)}^{-1}$ is an integer, $\gamma^{}_{L(R)}(\varepsilon,T)$
has a simple expression.
For example,
\begin{equation}
\gamma^{}_L(\varepsilon,T)=\cases{\displaystyle
\frac{\Gamma_L}{2T\cosh(\varepsilon/2T)},
&$g^{}_L=1$,\cr&\cr
\displaystyle
\frac{\varepsilon\Gamma_L}{4\Lambda T\sinh(\varepsilon/2T)},
&$g^{}_L=\frac{1}{2}$,\cr&\cr
\displaystyle
\frac{\Gamma_L(\pi^2T^2+\varepsilon^2)}
     {16\Lambda^2T\cosh(\varepsilon/2T)},
&$g^{}_L=\frac{1}{3}$,
}
\end{equation}
where we have used the simplified notation $\Gamma_{L}=\Gamma_{L00}$.
Thus, Eq.~(\ref{G_sequential}) reduces to Eq.~(\ref{G_Fermi}) when
$g^{}_L=g^{}_R=1$. 
It is interesting to observe that when $g^{}_L=g^{}_R=1/2$
Eq.~(\ref{G_sequential}) reduces to
\begin{equation}
G=
\frac{e^2}{4\hbar\Lambda}\frac{\Gamma_L\Gamma_R}{\Gamma_L+\Gamma_R}
\frac{\varepsilon/T}{\sinh(\varepsilon/T)},
\label{G_at_g=1/2}
\end{equation}
which has the same $\varepsilon$- and $T$-dependence as the
conductance of a quantum dot coupled to Fermi-liquid leads ($g=1$) at
$\Delta\ll T\ll E_C$ obtained by Glazman and Shekhter.\cite{GlazShek}
This coincidence occurred because of the effective halving of $g$ in
this temperature regime,\cite{FuruNaga} which we will discuss in the
next section.
It is also interesting to see that Matveev {\it et
al.}\cite{MatvGlazBara} have derived a formula similar to
Eq.~(\ref{G_sequential}) for the conductance of double quantum dots,
although the parameter $g$ has completely different physical meaning
in their case. 
Figure \ref{fig:g_s} shows the line shape of the linear conductance, 
Eq.~(\ref{G_sequential}), for the symmetric case, $\Gamma_L=\Gamma_R$
and $g^{}_L=g^{}_R=1$, 1/2, and 1/3.
The conductance is normalized by the peak conductance of the $g=1$
case, $G_{\max}=e^2\Gamma/8\hbar T$.
For fixed $\Gamma$ and $\varepsilon$, $G$ becomes smaller with
decreasing $g$. 
If scaled properly, however, the three curves in Fig.~\ref{fig:g_s}
can be made very similar to each other.
This is because for any $g$ the width of the peaks is proportional to
$T$ and the conductance decays exponentially for large
$|\varepsilon|$.
For example, the conductance of the $g=\frac12$ case
(\ref{G_at_g=1/2}) is almost proportional to that of the $g=1$ case
(\ref{G_Fermi}) with $T\to1.25T$.\cite{Beenakker}
It is the temperature dependence of the peak conductance,
$G(\varepsilon=0)\propto T^{\frac{1}{g}-2}$, that changes
qualitatively.
We hope this can be tested experimentally in the near future.
The anomalous temperature dependence of the peak conductance is a 
signature of the power-law decay of the propagators in the TL 
liquids, Eqs.~(\ref{G_A}) and (\ref{G_D}).

When the left and right leads are different TL liquids ($g^{}_L\ne
g^{}_R$), the peak conductance is roughly given by
$\min\!\left[
\frac{T}{\Gamma_L}\left(\frac{T}{\Lambda}\right)^{1-\frac{1}{g^{}_L}},
\frac{T}{\Gamma_R}\left(\frac{T}{\Lambda}\right)^{1-\frac{1}{g^{}_R}}
\right]$.
If $g^{}_L<1/2$ and $g^{}_R>1/2$, and
$\gamma^{}_R(0,T)\ll\gamma^{}_L(0,T)\ll1$ at some temperature
$(T<\Delta)$, then
the conductance may have a nonmonotonic temperature dependence.
With lowering temperature, the peak conductance first increases
as $G\propto T^{\frac{1}{g^{}_R}-2}$ and then decreases as
$G\propto T^{\frac{1}{g^{}_L}-2}$.
This nonmonotonic temperature dependence would be observed, for
example, for the sequential tunneling between the $\nu=1$ edge state
and the $\nu=1/3$ edge state.
When both $g^{}_L$ and $g^{}_R$ are larger (smaller) than $1/2$, on
the other hand, the conductance should monotonically increase 
(decrease) with decreasing temperature.

Next we examine when the sequential-tunneling approximation is
valid for temperatures $T\ll\Delta$.
We find two conditions to be satisfied.
First, the conductance calculated perturbatively in the tunneling
matrix elements must be much smaller than the conductance quantum,
$e^2/h$. 
This leads to the condition
$\gamma^{}_L(0,T)\ll1$ and $\gamma^{}_R(0,T)\ll1$,
or equivalently,
\begin{equation}
T\ll
\Lambda\left(\frac{\Lambda}{\Gamma_{L,R}}\right)
  ^{g^{}_{L,R}/(1-2g^{}_{L,R})}
\label{cond_1}
\end{equation}
when $g_{L,R}<1/2$ and
\begin{equation}
T\gg
\Lambda\left(\frac{\Gamma_{L,R}}{\Lambda}\right)
  ^{g^{}_{L,R}/(2g^{}_{L,R}-1)}
\label{cond_2}
\end{equation}
when $g_{L,R}>1/2$.
For the Fermi-liquid leads ($g=1$) this condition reduces to
$T\gg\Gamma$.
Note that, when $g<1/2$, the condition (\ref{cond_1}) is satisfied
down to zero temperature, once it is valid at some high
temperature.
Second, the current carried by virtual tunneling processes must be
negligibly small compared with the contribution from the sequential
tunneling processes we have calculated.
For $\varepsilon\gg T$ the second-order perturbation in $H_T$ yields
the operator for the virtual tunneling,
\begin{equation}
H_{\rm vt}=
\frac{t_Lt_R}{\varepsilon}
\left[
\psi^\dagger({\rm A})\psi({\rm B})
\psi^\dagger({\rm C})\psi({\rm D})
+{\rm h.c.}\right].
\label{H_virtual}
\end{equation}
In lowest order, the probability that an electron virtually tunnels
from the left lead to the right lead is
\end{multicols}\widetext
\vspace{-6mm}\noindent\underline{\hspace{90mm}}
\begin{eqnarray}
P_{\rm vt}(L\to R)&=&
\left(\frac{t^{}_Lt^{}_R}{\hbar\varepsilon}\right)^2
\!\int^\infty_{-\infty}\!dt\,
e^{-ieVt/\hbar}
\langle
\psi^\dagger({\rm A},t)\psi({\rm A},0)
\rangle_L
\langle
\psi({\rm B},t)\psi^\dagger({\rm C},t)
\psi^\dagger({\rm B},0)\psi({\rm C},0)
\rangle_D
\nonumber\\
&&\hspace*{4cm}\times
\langle
\psi({\rm D},t)\psi^\dagger({\rm D},0)
\rangle_R.
\label{P_vt}
\end{eqnarray}
We may neglect the time-dependence of the two-particle propagator in
the dot for $T\ll\Delta$, and write 
$\langle
\psi^\dagger({\rm B},t)\psi({\rm C},t)
\psi({\rm B},0)\psi^\dagger({\rm C},0)
\rangle_D
=c_2$.
Here $c_2$ is a dimensionless constant which may depend on the
geometry and the mean free path of the dot.\cite{AverinNazarov}
From Eqs.~(\ref{G_A}), (\ref{G_D}), and (\ref{P_vt}), we get
\begin{equation}
P_{\rm vt}(L\to R)=
\frac{c_3\Gamma_L\Gamma_R\Lambda}{4\pi^2\hbar\varepsilon^2}
e^{-eV/2T}
\left(\frac{\pi T}{\Lambda}\right)^{\frac{1}{g_L}+\frac{1}{g_R}-1}
\frac{\left|\Gamma\!\left(
             {\frac{1}{2g_L}+\frac{1}{2g_R}+i\frac{eV}{2\pi T}}
             \right)\right|^2}
     {\Gamma\!\left(\frac{1}{g_L}+\frac{1}{g_R}\right)},
\end{equation}
where $c_3$ is a dimensionless constant of the same order as $c_2$.
The probability of the reverse process is
$P_{\rm vt}(R\to L)=e^{eV/T}P_{\rm vt}(L\to R)$.
Hence the linear conductance due to the virtual tunneling is
\begin{equation}
G_{\rm vt}=
\lim_{V\to0}\frac{-e}{V}[P_{\rm vt}(L\to R)-P_{\rm vt}(R\to L)]
=
\frac{c_3e^2\Gamma_L\Gamma_R}{4\pi\hbar\varepsilon^2}
\frac{\left|\Gamma\!\left(\frac{1}{2g_L}+\frac{1}{2g_R}\right)\right|^2}
     {\Gamma\!\left(\frac{1}{g_L}+\frac{1}{g_R}\right)}
\left(\frac{\pi T}{\Lambda}\right)^{\frac{1}{g_L}+\frac{1}{g_R}-2}.
\label{G_vt}
\end{equation}
This contribution has the same temperature dependence
$T^{\frac{1}{g^{}_L}+\frac{1}{g^{}_R}-2}$ as that of the tunneling
between TL liquids coupled by a single tunnel barrier.\cite{KaneFish}
In the case of Fermi-liquid leads ($g^{}_L=g^{}_R=1$), $G_{\rm vt}$ is
independent of temperature, and the virtual tunneling process
contributing to $G_{\rm vt}$ is called the elastic
cotunneling.\cite{AverinNazarov} 
When $\varepsilon\lesssim T$, the temperature $T$ serves as a lower
cutoff so that $\varepsilon^2$ in the denominator of the rhs of
Eq.~(\ref{G_vt})  should be replaced
with $(|\varepsilon|+2\pi T)^2$.
When $|\varepsilon|\ll T$, the conductance $G$,
Eq.~(\ref{G_sequential}), is much larger than $G_{\rm vt}$ if
Eq.~(\ref{cond_1}) or (\ref{cond_2}) is satisfied.
On the other hand, using the relation
$|\Gamma(\frac{1}{2g}+i\frac{\varepsilon}{2\pi T})|^2\approx
 2\pi(\varepsilon/2\pi T)^{\frac{1}{g}-1}e^{-\varepsilon/2T}$ for
$\varepsilon\gg T$, 
we find that the condition for $G\gg G_{\rm vt}$ at $\varepsilon\gg T$
is equivalent to 
\begin{equation}
\frac{e^{-\varepsilon/T}}{T}
\left[
\frac{1}{\Gamma_L}
\left(\frac{\varepsilon}{\Lambda}\right)^{1-\frac{1}{g_L}}
+
\frac{1}{\Gamma_R}
\left(\frac{\varepsilon}{\Lambda}\right)^{1-\frac{1}{g_R}}
\right]^{-1}
\gg
\frac{\Gamma_L\Gamma_R}{\varepsilon^2}
\left(\frac{\pi T}{\Lambda}\right)^{\frac{1}{g_L}+\frac{1}{g_R}-2}.
\label{cond_3}
\end{equation}
For $g_L=g_R=g$ this condition is simplified to
\begin{equation}
\varepsilon\lesssim
T\ln\!\left[\frac{T}{\Gamma_{L,R}}
            \left(\frac{\Lambda}{T}\right)^{\frac{1}{g}-1}\right],
\end{equation}
in which the argument of logarithm is much larger than 1 if the
condition $\gamma_{L,R}(0,T)\ll1$ is satisfied.

\begin{figure}
\begin{center}\leavevmode\epsfxsize=65mm \epsfbox{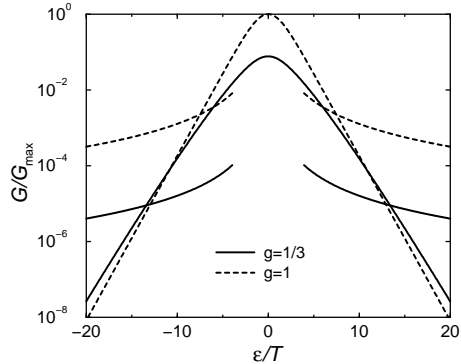}\end{center}
\caption{Line shape of a conductance peak.  Both the contribution from 
the sequential tunneling, Eq.~(\ref{G_sequential}), and that of the
virtual tunneling process, Eq.~(\ref{G_vt}), are shown for
$g^{}_L=g^{}_R=g=1$ and $1/3$.  The conductance is normalized 
by the peak value of the $g=1$ case, $G_{\max}=e^2\Gamma/8\hbar T$.
We have set $T/\Lambda=0.25$, $\Gamma/T=0.2$, and $c_3=1$.} 
\label{fig:g_v}
\end{figure}

From these considerations we conclude that, when Eq.~(\ref{cond_1}) or 
(\ref{cond_2}) is satisfied, the line shape of conductance peaks is
described by Eq.~(\ref{G_sequential}) around a peak and by
Eq.~(\ref{G_vt}) away from the peak, see Fig.~\ref{fig:g_v}.

Finally we briefly comment on the Kondo effect in the resonant
tunneling through a very small quantum dot or an impurity level.
In the Fermi-liquid case ($g^{}_L=g^{}_R=1$) it has been
shown\cite{Ng,Raikh} using the Anderson model that the conductance due
to tunneling processes via a virtual intermediate state 
logarithmically increases with lowering temperature and eventually
approaches $2e^2/h$ in the zero-temperature limit.
In the TL-liquid case ($g^{}_L,g^{}_R<1$) this kind of Kondo effect
does not happen for $|\varepsilon|\gtrsim T$, because
$G_{\rm vt}\propto T^{\frac{1}{g^{}_L}+\frac{1}{g^{}_R}-2}$.
In other words the virtual tunneling is irrelevant in the
renormalization-group sense, in contrast to the Fermi-liquid case
where the virtual tunneling is marginally relevant.
Nevertheless there exists an analogue of the Kondo effect in the TL
liquid case when $g^{}_L=g^{}_R=1/2$.
In this case, as we saw in Eq.~(\ref{G_at_g=1/2}), the peak
conductance $G(\varepsilon=0)$ is independent of temperature in the
lowest-order calculation.
In fact, one can show\cite{KaneFish,Matveev} that on resonance
($\varepsilon=0$) the tunneling is marginally relevant so that the
peak conductance increases logarithmically with lowering temperature
when higher-order terms are included.
It is, however, important to note that our problem is not exactly the
same as the (multi-channel) Kondo effect due to the interference
between the tunneling processes through the left and right barriers,
unlike the situation ($\Delta\ll T\ll E_C$) discussed by
Matveev.\cite{Matveev}

\section{High-temperature regime ($\Delta\ll T\ll E_C$)}
\label{sec:high-T}

Let us next consider the high-temperature case where
$\Delta\ll T\ll E_C$.
We note that, although we can only expect $\Delta\lesssim E_C$ in
general, the condition $\Delta\ll E_C$ can be satisfied in some
relatively large quantum dots.
In this temperature regime we may still assume that the number of
electrons in the dot is either $N_0$ or $N_0+1$.
However, the electrons in the dot no longer stay in the lowest-energy
state but occupy excited states.
This gives time dependence to the propagators in the dot.
At $T\gg\Delta$ we can regard the discrete energy levels as
continuum,\cite{GlazShek} and the propagators in the dot may be
written as 
\begin{mathletters}
\begin{eqnarray}
\langle
\psi^\dagger({\rm B},t)\psi({\rm B},0)
\rangle_D
&=&
\langle
\psi({\rm B},t)\psi^\dagger({\rm B},0)
\rangle_D
=
\rho_B
\left\{
\frac{i\Delta}{\pi T}
\sinh\!\left[\frac{\pi T(t-i\delta)}{\hbar}\right]
\right\}^{-\frac{1}{g^{}_D}},
\label{G_B}\\
\langle
\psi^\dagger({\rm C},t)\psi({\rm C},0)
\rangle_D
&=&
\langle
\psi({\rm C},t)\psi^\dagger({\rm C},0)
\rangle_D
=
\rho_C
\left\{
\frac{i\Delta}{\pi T}
\sinh\!\left[\frac{\pi T(t-i\delta)}{\hbar}\right]
\right\}^{-\frac{1}{g^{}_D}},
\label{G_C}
\end{eqnarray}
\end{mathletters}\noindent
where $\rho_B$ and $\rho_C$ are electron densities at the point B and
C.
Without a magnetic field [Fig.~\ref{fig:dots}(a)] the exponent
$g^{}_D$ is 1 ('Fermi liquid'), whereas $g^{}_D$ equals the filling
factor $\nu^{}_D=1/(2n+1)$ if the dot
is in the fractional quantum Hall regime [Fig.~\ref{fig:dots}(b)].
Using Eqs.~(\ref{G_A}), (\ref{G_D}), (\ref{G_B}), and (\ref{G_C}),
we can repeat the calculation in the previous section to derive the
linear conductance in the sequential tunneling approximation.
Since the excited states are already taken into account in
Eqs.~(\ref{G_B}) and (\ref{G_C}), we can simply use the result for the 
single-level case ($T\ll\Delta$), Eq.~(\ref{G_sequential}), with
appropriate modification.
We thus find that the conductance is given by
\begin{equation}
G=
\frac{e^2}{2\hbar\cosh(\varepsilon/2T)}
\frac{\tilde\gamma^{}_L(\varepsilon,T)\tilde\gamma^{}_R(\varepsilon,T)}
    {\tilde\gamma^{}_L(\varepsilon,T)+\tilde\gamma^{}_R(\varepsilon,T)}
\label{G_sequential_high}
\end{equation}
with
\begin{equation}
\tilde\gamma^{}_{L(R)}(\varepsilon,T)=
\frac{\widetilde\Gamma_{L(R)}}{2\pi T}
\left(\frac{\pi T}{\Lambda}\right)^{\frac{1}{g^{}_{L(R)}}-1}
\left(\frac{\pi T}{\Delta}\right)^{\frac{1}{g^{}_D}}
\frac{\left|\Gamma\!\left(\frac{1}{2g^{}_{L(R)}}+\frac{1}{2g^{}_D}
                         +i\frac{\varepsilon}{2\pi T}\right)\right|^2} 
     {\Gamma\!\left(\frac{1}{g^{}_{L(R)}}+\frac{1}{g^{}_D}\right)}.
\label{tilde_gamma}
\end{equation}
Here $\widetilde\Gamma_L=2\pi c^{}_At^2_L\rho_B/\alpha\Lambda$ and
$\widetilde\Gamma_R=2\pi c^{}_Dt^2_R\rho_C/\alpha\Lambda$.
The parameters $g^{}_{L}$ and $g^{}_R$ are effectively changed into
$g^{}_L\to\tilde g^{}_L\equiv g^{}_Lg^{}_D/(g^{}_L+g^{}_D)$
and
$g^{}_R\to\tilde g^{}_R\equiv g^{}_Rg^{}_D/(g^{}_R+g^{}_D)$.
Note that, when $g^{}_{L(R)}=g^{}_D$,
the parameter $g_{L(R)}$ is effectively halved:\cite{FuruNaga}
$\tilde g^{}_{L(R)}=g^{}_{L(R)}/2$.
This is the reason why Eq.~(\ref{G_sequential}) with
$g^{}_L=g^{}_R=1/2$ reproduced the high-temperature conductance of a
quantum dot coupled to Fermi-liquid leads
($g^{}_L=g^{}_R=1$).\cite{GlazShek}

The same exponent $\tilde g^{}_{L(R)}$ also appears in the conductance 
due to the virtual tunneling process.
This virtual tunneling is known as the inelastic cotunneling in the
Fermi liquid case.\cite{AverinNazarov}
For $T\gg\Delta$ an electron tunneling through the left tunnel
barrier may be different from an electron tunneling through the right
barrier, so that we may approximate the two-particle propagator as
$\langle
\psi({\rm B},t)\psi^\dagger({\rm C},t)
\psi^\dagger({\rm B},0)\psi({\rm C},0)
\rangle_D\approx
\langle
\psi({\rm B},t)\psi^\dagger({\rm B},0)
\rangle_D
\langle
\psi^\dagger({\rm C},t)\psi({\rm C},0)
\rangle_D$.
From Eqs.~(\ref{G_A}), (\ref{G_D}), (\ref{P_vt}), (\ref{G_B}), and
(\ref{G_C}), we get
\begin{equation}
G_{\rm vt}=
\frac{e^2\widetilde\Gamma_L\widetilde\Gamma_R}{8\pi\hbar\varepsilon}
\frac{\left|\Gamma\!\left(\frac{1}{2g^{}_L}+\frac{1}{2g^{}_R}
                          +\frac{1}{g^{}_D}\right)\right|^2}
     {\Gamma\!\left(\frac{1}{g^{}_L}+\frac{1}{g^{}_R}
                    +\frac{2}{g^{}_D}\right)}
\left(\frac{\pi T}{\Lambda}\right)
    ^{\frac{1}{g^{}_L}+\frac{1}{g^{}_R}-2}
\left(\frac{\pi T}{\Delta}\right)^{\frac{2}{g^{}_D}}.
\end{equation}
\begin{multicols}{2}\noindent
The conductance is proportional to
$T^{\frac{1}{g^{}_L}+\frac{1}{g^{}_R}+\frac{2}{g^{}_D}-2}$.
For the Fermi-liquid case ($g^{}_L=g^{}_R=g^{}_D=1$) this reduces to
$G_{\rm vt}\propto T^2$, in agreement with the inelastic
cotunneling theory.\cite{AverinNazarov}

\section{Conclusions}
\label{sec:conclusion}

In this paper we have studied the resonant tunneling through a quantum 
dot coupled to TL liquids in the weak-tunneling limit.
We have considered both the sequential tunneling process and the
tunneling process via a virtual intermediate state (cotunneling) to
calculate the linear conductance at temperatures $T\ll E_C$.
Within this approximation we have determined the line shape of the
conductance peaks as a function of a gate voltage.
At $T\lesssim\Delta$ the peak height and width are proportional to
$T^{\frac{1}{g}-2}$ and $T$, respectively.
This approach is justified in the weak-tunneling limit where the
conductance is much smaller than $e^2/h$.
In contrast to the Fermi-liquid case ($g=1$) where the approximation
breaks down at $T<\Gamma$, our result [Eqs.~(\ref{G_sequential}) and
(\ref{G_vt})] is valid down to zero temperature when the TL-liquid
parameters $g^{}_L$ and $g^{}_R$ are smaller than 1/2.

The edge states of the fractional quantum Hall liquids with filling
factor $\nu=1/(2m+1)$ correspond to the case $g=\nu$.
Hence the tunneling through a quantum dot weakly coupled to the edge
states is described by our theory in the whole 
temperature range.
We hope that the theory can be tested experimentally in the near
future.
The anomalous temperature dependence of the peak height
$T^{\frac{1}{\nu}-2}$ and careful fitting of the line shape to
Eq.~(\ref{G_sequential}) will give another firm evidence for the TL
liquid behavior of the edge states.
The anomalous exponent $\frac{1}{\nu}-2$ is a direct consequence of
the power-law tunnel density of states $\rho(E)\propto
E^{\frac{1}{\nu}-1}$ in TL leads and of the discrete energy
spectrum $\rho(E)\propto\delta(E)$ in a quantum dot.
It is also interesting to note that, when a quantum dot is weakly
coupled to the $\nu=1$ edge at one tunneling contact and to the
$\nu=1/3$ edge states at the other, the linear conductance may exhibit
a nonmonotonic temperature dependence.

We note that there are some cases in which the
sequential-tunneling picture is not applicable even when the bare
tunneling matrix elements are small.
For example, if $g^{}_L$ and $g^{}_R$ are larger 1/2,
the tunneling rates through the left and right tunnel barrier grow
with decreasing temperature. 
This means that the transport through a quantum dot becomes
coherent and the sequential-tunneling approximation breaks down at low 
temperature where $\gamma(0,T)\gtrsim1$. 
This coherent transport in the low-temperature limit is described
better starting from small-barrier (strong-tunneling) limit.
In this limit it was shown by Kane and Fisher\cite{KaneFish} that in
the symmetric case ($g_L=g_R>1/2$ and $t_L=t_R$) the backward
scattering is renormalized to zero when $\varepsilon=0$ and that the
line shape of conductance peaks approaches a universal
form\cite{KaneFish,Moon,Fendley,note3} 
below a crossover temperature at which $\gamma(0,T)\approx1$.
This is also the case for systems with $g^{}_L=g^{}_R=1/2$, in which
the tunneling at $\varepsilon=0$ is marginally relevant\cite{KaneFish}
and increases logarithmically with decreasing
temperature.\cite{Matveev} 
We emphasize again, however, that for the resonant tunneling between
the edge states in the fractional quantum Hall liquids as well as
between the 1D quantum wires with sufficiently strong repulsive
interactions, the sequential-tunneling approach developed in this
paper gives the correct description even in the low-temperature
limit.

\acknowledgements
The author would like to thank A.~MacDonald and
N.~Nagaosa for useful discussions
and M.~Sigrist for helpful comments on the manuscript.

\end{multicols}


\begin{references}
\bibitem{note_Chamon} Chamon and Wen\cite{Chamon} used a method which
is equivalent to the master-equation approach to calculate the current
through a single impurity level.
In some sense the present paper can be viewed as a generalization of
Ref.~\onlinecite{Chamon} to the case where there are many impurity
levels.
\bibitem{note1} In these quantum Hall states the electrons are spin
polarized, so that we can neglect spin indices.
\bibitem{note} The particles tunneling between the quantum dot and a
quantum Hall edge state are electrons, not the Laughlin
quasiparticles.
\bibitem{note2}
In Ref.~\onlinecite{FuruNaga} Nagaosa and the present author
incorrectly predicted a slightly different line shape, though they
obtained the correct temperature dependence of the peak conductance,
which was the main issue in Ref.~\onlinecite{FuruNaga}.
This error is made because $T\gamma(0,T)$ is used as a level width of
an intermediate state in Ref.~\onlinecite{FuruNaga}.
If we replace $\Gamma$ in Eq.~(11) of Ref.~\onlinecite{FuruNaga} with
$T\cosh(\varepsilon/2T)\gamma(\varepsilon,T)$, we get the correct
result, Eq.~(\ref{G_sequential}) of the present paper.
The calculation in Ref.~\onlinecite{FuruNaga} corresponds to taking $T$
as a low-energy cutoff for the renormalization even when
$\varepsilon>T$.
Note that
$T\cosh(\varepsilon/2T)\gamma(\varepsilon,T)
\propto(\max\{2\pi T,\varepsilon\})^{\frac{1}{g}-1}$.
\bibitem{note3} The universal line shape was derived in
Refs.~\onlinecite{KaneFish} by integrating out a massive
mode to reduce a double-barrier system to a single-barrier problem.
This procedure is justified only in the limit of weak backscattering.
\end{references}
\end{document}